\documentclass[aps,prd,preprint,groupedaddress]{revtex4}
\usepackage{graphicx}

 \newcommand{\nc}{\newcommand}
 \nc{\mb}[1]{\makebox[#1]{}}
 \nc{\CC}{{\scriptscriptstyle CC}}
 \nc{\NC}{{\scriptscriptstyle NC}}
 \nc{\V}{{\rm v}}
 \nc{\W}{{\scriptscriptstyle W}}
 \nc{\X}{{\scriptscriptstyle X}}
 \nc{\Z}{{\scriptscriptstyle Z}}
 \nc{\CS}{{\scriptscriptstyle CS}}
 \nc{\DY}{{\scriptscriptstyle DY}}
 \nc{\PW}{{\scriptscriptstyle PW}}
 \nc{\SB}{{\scriptscriptstyle SB}}
 \nc{\CSV}{{\scriptscriptstyle CSV}}
 \nc{\GLS}{{\scriptscriptstyle GLS}}
 \nc{\CIB}{{\scriptscriptstyle CIB}}
 \nc{\PT}{{\scriptscriptstyle PT}}
 \nc{\IE}{{\it i.e.,\ }}
 \nc{\EG}{{\it e.g.,\ }}
 \nc{\EA}{{\it et al.\ }}
 \nc{\AH}{{\it ad hoc\ }}
 \nc{\CHPT}{{$\chi_{\PT}$\ }}

\nc{\NCA}{{\em Nuovo Cimento}}
\nc{\NIM}{{\em Nucl. Instrum. Methods}}
\nc{\NIMA}{{\em Nucl. Instrum. Methods} A}
\nc{\NPB}{{\em Nucl. Phys.} B}
\nc{\PLB}{{\em Phys. Lett.}  B}
\nc{\PRL}{{\em Phys. Rev. Lett.}}

\nc{\PRD}{{\em Phys. Rev.} D}
\nc{\PRC}{{\em Phys. Rev.} C}
\nc{\ZPC}{{\em Z. Phys.} C}

\nc{\st}{\scriptstyle}
\nc{\sst}{\scriptscriptstyle}
\nc{\mco}{\multicolumn}
\nc{\epp}{\epsilon^{\prime}}
\nc{\vep}{\varepsilon}
\nc{\ra}{\rightarrow}
\nc{\ppg}{\pi^+\pi^-\gamma}
\nc{\nuN}{{\nu N_0}}
\nc{\nubN}{{\overline{\nu} N_0}}
\nc{\snuNC}{{\langle \sigma^{\nuN}_{\NC}\rangle }}
\nc{\snubNC}{{\langle \sigma^{\nubN}_{\NC}\rangle }}
\nc{\snuCC}{{\langle \sigma^{\nuN}_{\CC}\rangle }}

\nc{\snubCC}{{\langle \sigma^{\nubN}_{\CC}\rangle }}
\nc{\Rnu}{{R^{\nu}}}
\nc{\Rnub}{{R^{\overline{\nu}}}}
\nc{\sintW}{{\sin^2 \theta_{\W} }}
\nc{\MS}{{\overline{MS}}}
\nc{\vp}{{\bf p}}
\nc{\rz}{{1\over \rho_0^2}}
\nc{\ko}{K^0}
\nc{\kb}{\bar{K^0}}
\nc{\al}{\alpha}
\nc{\ab}{\bar{\alpha}}
\nc{\be}{\begin{equation}}
\nc{\ee}{\end{equation}}
\nc{\bea}{\begin{eqnarray}}
\nc{\eea}{\end{eqnarray}}

\begin{document}

\titlepage

\title{Charge Symmetry Violation Corrections to Determination of 
 \\ the Weinberg Angle in Neutrino Reactions} 

\author{J.T.Londergan}

\email{tlonderg@indiana.edu}
\affiliation{Department of Physics and Nuclear
            Theory Center,\\ Indiana University,\\ 
            Bloomington, IN 47405, USA}

\author{A.W.Thomas}
\email{athomas@physics.adelaide.edu.au}
\affiliation {Department of Physics and Mathematical Physics,\\ 
                and Special Research Center for the
                Subatomic Structure of Matter,\\ 
                University of Adelaide,
                Adelaide 5005, Australia}
\date{\today}


\begin{abstract} 
We show that the correction to the Paschos-Wolfenstein relation
associated with charge symmetry violation in the valence quark
distributions is essentially model independent. It is proportional 
to a ratio of quark momenta that 
is independent of $Q^2$. This result provides a natural explanation
of the surprisingly good agreement found between our earlier estimates
within several different models. 
When applied to the recent NuTeV measurement, this effect significantly
reduces the discrepancy with other determinations of the Weinberg angle.
\end{abstract}




\maketitle


In 1973, Paschos and Wolfenstein \cite{Pas73} derived an expression 
using the ratio of neutral-current and charge-changing neutrino 
interactions on 
isoscalar targets.  This ratio is 
\be
 R^- \equiv { \rz \left( \snuNC - \snubNC \right) \over 
 \snuCC - \snubCC } = {1\over 2} - \sintW  .
\label{eq:PasW} 
\ee  
In Eq.\ \ref{eq:PasW}, $\snuNC$ and $\snuCC$ are respectively the 
neutral-current and charged-current inclusive, total 
cross sections for neutrinos on an 
isoscalar target.  The quantity $\rho_0 \equiv M_{\W}/(M_{\Z}\,\cos 
\theta_{\W})$ is one in the Standard Model.  
{}From this ratio, one can obtain an independent measurement of the Weinberg 
angle ($\sintW$).  

The NuTeV group recently measured neutrino charged-current and 
neutral-current cross sections on iron \cite{NuTeV}.  From the ratios of
these cross sections for neutrinos and anti-neutrinos, 
they extracted 
$\sintW = 0.2277 \pm 0.0013 ~(stat) \pm 0.0009 ~(syst)$.  This 
value is three standard deviations above the measured fit to 
other electroweak processes, $\sintW = 0.2227 \pm 0.00037$ \cite{EM00}.  
The discrepancy between the NuTeV measurement and determination of the 
Weinberg angle from electromagnetic measurements is surprisingly large, 
and it may be evidence of physics beyond the 
Standard Model.   

As the NuTeV experiment did not strictly involve the Paschos-Wolfenstein
relation, Eq.\ (\ref{eq:PasW}), there are a number of additional
corrections that need to be considered, such as differences in shadowing
for photons, $W^{\pm}$ and $Z^0$'s \cite{Miller:2002xh}, 
asymmetries in $s$ and $\bar{s}$ distributions \cite{Cao:2003ny} and so on
-- Ref.\ \cite{Davidson:2001ji} summarizes corrections to the NuTeV 
result from within and outside the Standard Model.  
In addition, Eq.\ (\ref{eq:PasW}) is valid only for an isoscalar target 
and it is based upon the assumption of charge symmetry.  There is thus 
a premium on knowing the corrections as accurately as possible. 

Let us first review the corrections due to the fact that $N \ne Z$ for 
the iron target.  The corrections take the form \cite{Davidson:2001ji}
\be 
 \Delta R_{I} = -\left[ 3\Delta_u^2 + \Delta_d^2 + {4\alpha_s \over 9\pi}
 \left(\bar{g}_L^2 - \bar{g}_R^2 \right) \right] \,\left({N-Z \over A}
 \right) \,\left[{U_{\V} - D_{\V}  \over U_{\V} + D_{\V}} \right] 
\label{eq:isoscal} 
\ee
where 
\bea
 \Delta_q^2 &\equiv& \left(g_L^q \right)^2 - \left(g_R^q \right)^2~;  
 \hspace{0.5in}  3\Delta_u^2 + \Delta_d^2 = 1 - {7\over 3}\sintW ~; 
 \nonumber \\ \bar{g}_L^2 - \bar{g}_R^2 &=& \left(g_L^u \right)^2 + 
\left(g_L^d \right)^2 - \left(g_R^u \right)^2 - \left(g_R^d \right)^2
 = {1\over 2} - \sintW \nonumber \\
 Q_{\V} &\equiv& \int_0^1 \, x\,q_{\V}(x) \,dx  
\label{eq:deldef}
\eea
The additional QCD radiative term in Eq.\ \ref{eq:isoscal} was 
calculated by Davidson \EA, Ref.\ \cite{Davidson:2001ji}; it is 
quite small at the $Q^2$ for the NuTeV experiment.  The final 
term in Eq.\ \ref{eq:isoscal} involves the ratio of momentum carried 
by up and down valence quarks.  Since both numerator and 
denominator involve the same moments of QCD non-singlet 
parton distributions, they evolve identically, so this 
ratio can be evaluated at any convenient value of $Q^2$. 
Using the CTEQ3D parton distributions \cite{CTEQ3} in eq.\ 
\ref{eq:isoscal}, one 
obtains $\delta R_I = -0.0126$.  The NuTeV group 
has emphasized \cite{NuTeV,NuTeV2} that they do not 
actually measure the Paschos-Wolfenstein ratio, but instead combine 
separate measurements of ratios of neutral to charged-current cross 
sections for neutrinos and anti-neutrinos with a full 
Monte Carlo simulation of their experiment.  Using their simulation, 
the NuTeV group reports an isoscalar correction of $-0.0080$.  This 
represents a 36\% reduction from the Paschos-Wolfenstein correction, and 
the NuTeV group cited a very small error for this correction 
\cite{ZelPC}.  Kulagin \cite{Kul03} claimed that the uncertainty in 
this correction is likely to be considerably larger.  
The largest uncertainty in Eq.\ \ref{eq:isoscal} is the momentum 
carried by up and down valence quarks, and according to Davidson 
\EA \cite{Davidson:2001ji}, these quantities are known rather accurately.  

Davidson \EA \cite{Davidson:2001ji} noted that, although charge symmetry 
violating (CSV) corrections are likely to be small, these effects 
could in principle generate a substantial correction to the 
NuTeV result.  Recently, we calculated CSV contributions to the NuTeV 
experiment arising from the small 
difference of $u$ and $d$ quark masses \cite{Lon03}. 
{}Following earlier work on CSV in parton 
distributions \cite{Sat92,Rodionov:cg}, our
method involved calculating CSV distributions 
at a low momentum scale, and using QCD evolution to generate the CSV
distributions at the $Q^2$ 
values appropriate for the NuTeV experiment. 
We obtained a CSV correction to the 
NuTeV result $\Delta R_{\CSV} \sim -0.0015$.  The NuTeV group also 
reported an estimate of the CSV parton distributions, using a rather 
different procedure \cite{NuTeV2}; they obtained  
a much smaller correction than ours, $\Delta R_{\CSV} \sim +0.0001$. 
The large discrepancy between these two results suggested that the 
CSV correction might be strongly dependent on the 
starting scale, $Q_0^2$, the phenomenological valence parton distribution 
chosen, or other details of the calculation.

Here, we will demonstrate that one can obtain firm predictions for the 
CSV corrections, and that the CSV contributions to the 
Paschos-Wolfenstein ratio are essentially model independent.   
The charge symmetry violating contribution to the Paschos-Wolftenstein ratio 
has the form 
\be 
 \Delta R_{\CSV} = \left[ 3\Delta_u^2 + \Delta_d^2 + {4\alpha_s \over 9\pi}
 \left(\bar{g}_L^2 - \bar{g}_R^2 \right) \right] \,\left[ {\delta U_{\V} - 
 \delta D_{\V}  \over 2(U_{\V} + D_{\V}) }\right] 
\label{eq:CSV} 
\ee
where 
\bea 
 \delta Q_{\V} &=& \int_0^1 \, x\,\delta q_{\V}(x) \,dx \nonumber \\ 
 \delta d_{\V}(x) &=& d_{\V}^p(x)- u_{\V}^n(x)~; \hspace{1.0cm}
 \delta u_{\V}(x) = u_{\V}^p(x)- d_{\V}^n(x) ~.
 \label{eq:CSVadd} 
\eea
The denominator in the final term in Eq.\ (\ref{eq:CSV}) gives the total 
momentum carried by up and down valence quarks, while the numerator 
gives the charge symmetry violating momentum 
difference, \EG $\delta U_{\V}$ is the total momentum carried by up 
quarks in the proton minus the momentum of down quarks in the neutron.  
As for the isoscalar correction, this ratio is completely 
independent of $Q^2$ and can be evaluated at any convenient value of $Q^2$.

In our paper \cite{Lon03}, we used an analytic approximation to the charge 
symmetry violating valence parton distributions that was initially proposed 
by Sather \cite{Sat92}.  His equations were 
\bea 
 \delta d_{\V}(x) &=& -\frac{\delta M}{M} \frac{d}{dx} \left[ 
 x d_{\V}(x)\right] - \frac{\delta m}{M} \frac{d}{dx} d_{\V}(x) \nonumber \\
 \delta u_{\V}(x) &=& \frac{\delta M}{M} \left( - \frac{d}{dx}\left[ 
 x u_{\V}(x)\right] + \frac{d}{dx} u_{\V}(x) \right) 
\label{eq:Satanl}
\eea   
In Eq.\ (\ref{eq:Satanl}), $M$ is the average nucleon mass, $\delta M = 1.3$ 
MeV is the neutron-proton mass difference, and $\delta m = m_d - m_u \sim 
4$ MeV is the down-up quark mass difference. Eq.\ (\ref{eq:Satanl}) is 
valid for a low scale, $Q_0^2$, appropriate 
for a (valence dominated) quark or bag model~\cite{Signal:yc}.   

Sather's approximation allows us to evaluate directly the relevant 
integrals of the CSV distributions.  For $\delta D_{\V}$, we obtain  
\bea 
 \delta D_{\V} &=& \int_0^1 \, x  \left[ -\frac{\delta M}{M} \frac{d}{dx} 
 (x d_{\V}(x)) - \frac{\delta m}{M} \frac{d}{dx} d_{\V}(x) \right] 
 \, dx \nonumber \\ &=& \frac{\delta M}{M} \int_0^1 \, x \, d_{\V}(x)\, dx + 
 \frac{\delta m}{M} \int_0^1 \, d_{\V}(x)\, dx = \frac{\delta M}{M} D_{\V} 
 + \frac{\delta m}{M} 
\label{eq:intDv}
\eea   
The second line of Eq.\ (\ref{eq:intDv}) is obtained by integrating by parts, 
using the fact that there is one down valence quark in the proton.  In 
analogous fashion, the integral of the up quark CSV distribution is 
\bea 
 \delta U_{\V} &=& \frac{\delta M}{M} \, \int_0^1 \, x\, \left( - 
 \frac{d}{dx}\left[ x u_{\V}(x)\right] + \frac{d}{dx} u_{\V}(x) \right) 
 \, dx  \nonumber \\
 &=& \frac{\delta M}{M} \left( \int_0^1 \,x \, u_{\V}(x)\, dx - \int_0^1 \,  
 u_{\V}(x)\, dx \right) = \frac{\delta M}{M} \left( U_{\V} - 2 \right) 
\label{eq:intUv}
\eea  

Using Sather's approximation relating CSV distributions to valence 
quark distributions, Eqs.\ (\ref{eq:intDv}) and (\ref{eq:intUv}) show that 
the CSV correction to the Paschos-Wolfenstein ratio depends only on 
the fraction of the nucleon momentum carried by up and down valence quarks.  
At no point do we have to calculate specific CSV distributions.  
At the bag model scale, $Q_0^2 \approx 0.5$ GeV$^2$, the momentum fraction 
carried by down valence quarks,   
$D_{\V}$, is between $0.2-0.33$, and the total momentum fraction carried 
by valence quarks is $U_{\V} + D_{\V} \sim .80$.  From Eqs. (\ref{eq:intDv}) 
and (\ref{eq:intUv}) this gives $\delta D_{\V} \approx 0.00463$, 
$\delta U_{\V} \approx -0.00203$.  
Consequently, evaluated at the quark model scale, the CSV correction to the 
Paschos-Wolfenstein ratio is 
\be 
 \Delta R_{\CSV} \approx \left[ 3\Delta_u^2 + \Delta_d^2 \right] 
 \,{\delta U_{\V} - \delta D_{\V} \over 2(U_{\V} + D_{\V}) } \approx 
 -0.00203 .
\label{eq:CSVnut}
\ee
Once the CSV correction has been calculated at some quark model scale,
$Q^2_0$, the ratio appearing in Eq.~(\ref{eq:CSV}) is independent of
$Q^2$, because both the numerator and
denominator involve the same moment of non-singlet distributions 
(in Eq.\ \ref{eq:CSVnut} we have dropped the QCD radiative 
correction, which is very small at the $Q^2$ appropriate to the NuTeV 
measurements). 

We stress that both Eqs.~(\ref{eq:intDv}) and (\ref{eq:intUv}) are only
weakly dependent on the choice of quark model scale -- through the
momentum fractions $D_{\V}$ and $U_{\V}$, which are slowly varying functions
of $Q^2_0$, and are not the dominant terms in
those equations. This, together with the $Q^2$-independence of the
Paschos-Wolfenstein ratio, Eq.~(\ref{eq:CSV}), under QCD
evolution, explains why our previous results, obtained with  
different models at different $Q^2$ values~\cite{Lon03}, were so 
similar. For example, the  
result of Eq.\ (\ref{eq:CSVnut}) $\Delta R_{\CSV} = -0.00203$, 
at $Q_0^2 = 0.5$ GeV$^2$, is virtually identical with  
results using the Rodionov CSV distribution ($-0.0020$) and  
the Sather CSV distribution ($-0.0021$), at $Q^2 = 10$ and $12.6$ GeV$^2$, 
respectively. Using Eqs.~(\ref{eq:intDv}) and (\ref{eq:intUv}), we 
also calculated a CSV distribution using the CTEQ4LQ phenomenological parton 
distribution \cite{CTEQ4} at $Q^2 = 0.49$ GeV$^2$, evolved this to 
20 GeV$^2$, and obtained $\Delta R_{\CSV} = -0.0021$ \footnote{this 
value is not correct in Table II of our paper, but is  
corrected in the eprint \cite{Lon03}.}.   

Cao and Signal \cite{Cao:2003ny} point out some limitations of 
Sather's approximation, Eq.\ \ref{eq:Satanl}. However, we have 
compared $\delta U_{\V} - \delta D_{\V}$ obtained by Sather 
\cite{Sat92}, and by Rodionov \EA \cite{Rodionov:cg}, who did not 
use Sather's approximation, and they differ by only a 
few percent.     

As noted earlier, the NuTeV group \cite{NuTeV,NuTeV2} do not 
measure the Paschos-Wolfenstein ratio, but combine separate measurements 
of neutrinos and anti-neutrinos with a  
Monte Carlo simulation of their experiment.  They have    
produced functionals giving the sensitivity of their observables 
to various effects, including parton charge symmetry violation.  These are 
summarized in a single integral 
\be
  \Delta {\cal E} = \int_0^1 \, F\left[ {\cal E}, \delta; x\right] 
  \, \delta(x) \, dx .
\label{eq:Func}
\ee
Eq.\ (\ref{eq:Func}) gives the change in the extracted quantity 
${\cal E}$ resulting from the symmetry violating quantity $\delta(x)$.  
The functionals appropriate for the observable $\sintW$ and 
the parton CSV distributions, were provided in Ref.\ \cite{NuTeV2}.  

In our previous paper we found that including the NuTeV functional 
with evolved distributions decreased the CSV correction by about 33\% 
from the Paschos-Wolfenstein result.  This is very similar 
to the 36\% reduction obtained by NuTeV for the isoscalar correction.   
After applying this reduction, the CSV correction to the NuTeV experiment 
is $-0.0015$.  When the NuTeV measurement is adjusted accordingly, the 
disagreement between the NuTeV and electromagnetic results for   
$\sintW$ is reduced from $0.0050$ to $0.0035$ -- a 30\% decrease 
in that discrepancy.  

In conclusion, we have a robust prediction for the CSV contribution 
to the Paschos-Wolfenstein ratio, and also to the NuTeV measurement 
of the Weinberg angle.  The Sather approximation allows us to write integrals 
of $x \delta q_{\V}$ in terms of the total momentum carried by valence 
quarks. These integrals can be calculated without ever specifying 
the CSV distributions.  The Paschos-Wolfenstein ratio involves ratios of 
integrals that behave identically under QCD evolution, so these ratios 
are independent of $Q^2$. Despite the fact that parton charge symmetry 
violation has not been directly measured experimentally, and that 
parton CSV effects are predicted to be quite small, we have strong 
theoretical arguments regarding both the sign and magnitude of these 
corrections.  The CSV effects should make a significant contribution to 
the value for the Weinberg angle extracted from the NuTeV neutrino 
measurements.   

This work was supported in part by the Australian Research Council.  
One of the authors [JTL] was supported in part by National
Science Foundation research contract PHY--0070368.  The authors wish 
to thank G.P. Zeller and K. McFarland of the NuTeV collaboration,  
and W. Melnitchouk, for useful discussions regarding these issues.

\end{document}